\documentclass[twocolumn, unsortedaddress, superscriptaddress]{revtex4}
\usepackage{graphicx}
\usepackage{color}
\usepackage{amsmath}

\newcommand{\be}{\begin{equation}}
\newcommand{\ee}{\end{equation}}
\newcommand{\bea}{\begin{eqnarray}}
\newcommand{\eea}{\end{eqnarray}}

\begin{document}

\title{Generation of photovoltage in graphene on a femtosecond time scale through
efficient carrier heating}

\author{K.\ J.\ Tielrooij}\email[Correspondence to:
]{frank.koppens@icfo.eu, niek.vanhulst@icfo.eu, klaas-jan.tielrooij@icfo.eu}
\author{L.\ Piatkowski}\thanks{Equal contribution}
\author{M.\ Massicotte}\thanks{Equal contribution}
\author{A.\ Woessner} \affiliation{ICFO - Institut de Ci\`encies Fot\`oniques, Mediterranean Technology Park, Castelldefels (Barcelona) 08860, Spain}
\author{Q.\ Ma} \affiliation{Department of Physics, Massachusetts Institute of Technology, Cambridge, MA 02139, USA}
\author{Y.\ Lee}
\author{K.\ S.\ Myhro}
\author{C.\ N.\ Lau} \affiliation{Department of Physics and Astronomy, University of California, Riverside, CA 92521, USA}
\author{P.\ Jarillo-Herrero} \affiliation{Department of Physics, Massachusetts Institute of Technology, Cambridge, MA 02139, USA}
\author{N.\ F.\ van Hulst}\email[Correspondence to:
]{frank.koppens@icfo.eu, niek.vanhulst@icfo.eu, klaas-jan.tielrooij@icfo.eu} \affiliation{ICFO - Institut de Ci\`encies Fot\`oniques,
Mediterranean Technology Park, Castelldefels (Barcelona) 08860,
Spain}\affiliation{ICREA-Instituci\'o
Catalana de Recerca i Estudis Avan\c{c}ats, 08010 Barcelona, Spain}
\author{F.\ H.\ L.\ Koppens} \email[Correspondence to:
]{frank.koppens@icfo.eu, niek.vanhulst@icfo.eu, klaas-jan.tielrooij@icfo.eu} \affiliation{ICFO - Institut de Ci\`encies Fot\`oniques,
Mediterranean Technology Park, Castelldefels (Barcelona) 08860,
Spain}

\newpage

\begin{abstract}
\textbf{
Graphene is a promising material for ultrafast and broadband photodetection. Earlier studies addressed the general operation of graphene-based photo-thermoelectric devices, and the switching speed, which is limited by the charge carrier cooling time, on the order of picoseconds. However, the generation
of the photovoltage could occur at a much faster time scale, as it is associated with the carrier heating time. Here, we measure the photovoltage generation time and find it to be faster than 50 femtoseconds. As a proof-of-principle application of this ultrafast photodetector, we use graphene to directly measure, electrically, the pulse duration of a sub-50 femtosecond laser pulse. The observation that carrier heating is ultrafast suggests that energy from absorbed photons can be efficiently transferred to carrier heat. To study this, we examine the spectral response and find a constant spectral responsivity between 500 and 1500 nm. This is consistent with efficient electron heating. These results are promising for ultrafast femtosecond and broadband photodetector applications.
}
\end{abstract}

\maketitle

Photovoltage generation through the photo-thermoelectric (PTE) effect occurs when light is focused at the interface of monolayer and bilayer graphene, or at the interface between regions of graphene with different Fermi energies $E_F$ \cite{Wei2009, Xu2010, Song2011, Gabor2011, Freitag2013, Koppens2014}. In such graphene PTE devices -- which operate over a large spectral range \cite{Bonaccorso2010, Echtermeyer2014} that extends even into the far-infrared \cite{Cai2014} -- local heating of electrons by absorbed light, in combination with a difference in Seebeck coefficients between the two regions, gives rise to a photo-thermoelectric voltage $V_{\rm PTE} = (S_2 - S_1)(T_{\rm el} - T_0)$. Here $S_1$ and $S_2$ are the Seebeck coefficients of regions 1 and 2, respectively, $T_{\rm el}$ is the hot electron temperature after photoexcitation and electron heating, and $T_0$ the temperature of the electrode heat sinks. The performance of PTE graphene devices is intimately connected to the dynamics of the photoexcited electrons and holes, which have mainly been studied in graphene samples through ultrafast optical pump-probe measurements \cite{George2008, Breusing2011, Gierz2013, Tielrooij2013, Johannsen2013, Lui2010, Jensen2014, Brida2013}. As shown in Fig.\ 1a, the dynamics start with \textit{i)} photo-excitation and electron-hole pair generation; followed by \textit{ii)} electron heating through carrier-carrier scattering, in competition with lattice heating, which both take place on a sub-100 fs time scale; and finally \textit{iii)} electron cooling by thermal equilibration with the lattice, which takes place on a picosecond time scale. The effect of the picosecond cooling step \textit{iii)} on the switching speed of graphene devices has been studied using time-resolved photovoltage scanning experiments with $\sim$200 fs time resolution \cite{Sun2012, Urich2011, Graham2013}. These studies showed that the picosecond electron cooling time limits the intrinsic photo-switching rate of these devices to a few hundred GHz, because faster switching would reduce the switching contrast, as the system does not have time to return to the ground state. Indeed, GHz switching speeds have been demonstrated in graphene-based devices \cite{Gan2013, Xia2009, Schall2014, Mueller2010, Pospischil2013}.
\\

However, the most crucial aspects of the photo-thermoelectric response are captured by the heating dynamics, as electron heating corresponds to photovoltage generation.  Additionally, these dynamics determine the ultimate intrinsic carrier heating efficiency and the resulting spectral response. Here, we measure the photovoltage generation time with an unprecedented time resolution, and assess its effect on the heating efficiency through spectral responsivity measurements. In an ideal photo-thermoelectric detector, all the absorbed photon energy is transferred rapidly to electron heat (before energy is lost through other channels). In this case, doubling the photon energy would lead to doubling of the photovoltage (see Fig.\ 1b), and would result in a flat spectral responsivity $R_{\rm PC} = I_{\rm PC} / P_{\rm exc}$, where $I_{\rm PC}$ is the generated photocurrent and $P_{\rm exc}$ is the excitation power. In strong contrast, the spectral response of conventional semiconductor-based detectors is not flat at all, since it is determined by the band gap: Photons with an energy below the band gap are not absorbed, i.e.\ $R_{\rm PC}$ = 0, and the excess energy above the band gap typically does not lead to an additional photoresponse, i.e.\ a decreasing $R_{\rm PC}$ with photon energy \cite{Sze}.

\section*{Photovoltage generation on a femtosecond time scale}

In order to capture the time scale of photovoltage generation, we perform time-resolved photovoltage measurements with the highest time resolution to date of $\sim$30 fs. We achieve this by using a broadband Ti:Sapphire 85 MHz oscillator (center frequency 800 nm, bandwidth $>$100 nm) that creates $<$20 fs pulses, and a pulse shaper that corrects for any dispersion (and thus pulse stretching) that the pulses pick up on their way from the laser to the device (see Fig.\ 1c and Supp.\ Info for details). The concept of the experiment is illustrated in Fig.\ 1d. We use a pair of ultrashort laser pulses, where the two pulses are separated by a controllable delay time $t$ and focused onto a single spot on the device. The two pulses contain different spectral components to suppress coherent artifacts. We then record the photocurrent, averaged over a large number of pulse pairs, as a function of delay $t$. After absorption of the first pulse, the electron temperature $T_{\rm el}$ rises by $\Delta T_{\rm el}$, and then starts cooling down to $T_0$. If the second pulse arrives before $T_{\rm el}$ reaches $T_0$ (i.e.\ when the delay between the two pulses is shorter than the cooling time), the electron temperature rises again but by less than $\Delta T_{\rm el}$. This is because the electronic heat capacity $C_{\rm el}$ of a degenerate electron gas increases with electron temperature \cite{Kittel} and $\Delta T_{\rm el} = \int_{Q_0}^{Q_0 + \Delta Q} dQ / C_{\rm el}(T_{\rm el}) $, with $Q_0$ the heat in the system before photoexcitation and $\Delta Q$ the absorbed power from a laser pulse. Thus, the additional amount of generated photovoltage by the second pulse is lower than when the pulses would each contribute independently. As a result, the photovoltage as a function of delay time directly reflects the time dynamics of the electron temperature and therefore also the PTE induced photovoltage.
\\

We apply our femtosecond photovoltage sensing technique to a graphene $pn$-junction device, that consists of a bottom and top dual-gated graphene flake (``dual-gated device", see Supp.\ Info). A scanning photocurrent image is shown in Fig. 2a, where the top gate and bottom gate voltages are such that a $pn$-junction is formed at the edge of the top gate. Figure 2b shows the photocurrent with the laser focused at this position, as a function of back-gate and top-gate voltages. The multiple sign reversals indicate that the photovoltage is generated through the PTE effect \cite{Song2011, Gabor2011}. We show the photocurrent generated in this device as a function of $t$ in Fig.\ 2c, which clearly shows a dip around $t = 0$. In Fig.\ 2d-e, we show the normalized photocurrent dip $\Delta I_{\rm PC}$ for two combinations of gate voltages, both in the $pn$-regime. The decays on both sides of $t=0$ reflect the cooling dynamics with a picosecond time scale, as observed in Refs.\ \cite{Sun2012, Urich2011, Graham2013}. Around $t=0$ we notice that the photocurrent dip $\Delta I_{\rm PC}$ is remarkably sharp, which is only possible when the time resolution of the complete system (i.e.\ the laser pulses and the graphene photoresponse) is sufficiently high. Any decrease of time resolution, either because of longer pulses or due to a slower generation of the photovoltage in graphene, would lead to a broadening of the apex of the inverted v-shape.
\\

We use two approaches to quantitatively determine the time scale of photovoltage creation. In the first approach, we characterize the sharpness of the photocurrent dip around $t=0$ by taking the time-derivative. In the Supp.\ Info we show that in the case of exponential heating dynamics, taking the derivative directly recovers the heat dynamics: ${{dI_{\rm PC}} / {dt}} (t) \approx (1 - e^{-t/\tau_{\rm heat}})e^{-t/\tau_{\rm cool}}$, where $\tau_{\rm heat}$ and $\tau_{\rm cool}$ are the electron heating and electron cooling time scales, respectively. Figure 2f shows the experimentally obtained ${{dI_{\rm PC}} / {dt}} (t)$ together with the heating dynamics using $\tau_{\rm heat} = $ 80 fs, obtained from a best fit. To show that we can indeed resolve time scales shorter than the time scale that was accessible in earlier time-resolved studies \cite{Graham2013,Sun2012,Urich2011}, we also plot the dynamics with $\tau_{\rm heat} = 250$ fs. This slower heating time clearly does not fit the experimental data. In the second approach we develop a model for the photovoltage as a function of $t$, based on heating dynamics induced by the pulse-pair excitation and including nonlinear heating (see Methods). We show the data for the first (second) gate voltage combination in Fig.\ 2d(e), together with the modeled photocurrent change $\Delta I_{\rm PC} (t)$. We find excellent agreement between data and model for $\tau_{\rm heat} =$ 50 (80) fs and a cooling time of $\tau_{\rm cool} =$ 1.3 (1.5) ps. As an illustration we show the model for $\tau_{\rm heat} = 250$ fs, which is clearly in strong disagreement with the data. We thus conclude that for the $pn$-junction configuration, photovoltage generation occurs within 50 fs.
\\

We now put this capability of ultrafast photovoltage generation into the perspective of an application. The switching speed of graphene optoelectronic devices using the direct photoresponse is limited by the picosecond cooling time, as shown by earlier reports \cite{Xia2009, Urich2011}, and limited to a few hundred GHz. However, we envision femtosecond photosensing applications by exploiting the nonlinear heating response and combining graphene PTE devices with time-differential operation. Here, we provide one proof-of-principle demonstration: we use the graphene photodetector to measure in a direct electrical signal the duration of an ultrashort femtosecond laser pulse. We compare the derivative photocurrent signal ${{dI_{\rm PC}} / {dt}} (t)$ with the optical cross correlation signal that is measured by overlapping the two laser pulses in a second harmonic generation crystal and monitoring the second harmonic signal at 400 nm as a function of delay time (orange area in Fig.\ 2f). The agreement shows that our graphene PTE device is capable of measuring the pulse width of the laser down to time scales below 50 fs, in a direct electrical signal, and without the use of non-linear crystals. We note that this technique will work over a much broader spectral range (from the ultraviolet \cite{Tielrooij2013} to the terahertz \cite{Cai2014}) than techniques based on two-photon absorption in silicon photodiodes or frequency conversion in nonlinear crystals, and has a similar sensitivity.

\section*{Efficient photo-induced carrier heating}

Having established that carrier heating and PTE photovoltage generation occur on a femtosecond time scale, we now address how this step affects the energy conversion efficiency of graphene PTE devices. The main question is if the carrier heating is fast enough to out-compete energy loss processes, such as optical phonon emission (step \textit{ii)} in Fig.\ 1a). To this end, we study the photoresponse for a wide range of photon energies, from 0.8 (1500 nm) to 2.5 eV (500 nm). We use a laser source (quasi-CW (continuous wave), since the pulse duration of 20 ps is larger than the cooling time of $\sim$1 ps) with a controllable wavelength and a constant excitation power $P_{\rm exc}$. In Fig.\ 3a, we show the (external) responsivity $R_{\rm PC}$ for the dual-gated $pn$-junction device.  This spectral response is dominated by the strongly wavelength-dependent absorption spectrum of the device, which is the result of reflections at the oxide-silicon interface and the sub-wavelength oxide thickness (see Supp.\ Info). This effect is very similar to the enhanced reflection for certain combinations of wavelength and oxide thickness, which enhance the contrast of graphene in an optical microscope \cite{Blake2007}. Indeed, the calculated graphene absorption (using Lumerical FDTD Solutions software) agrees well with the wavelength-dependent responsivity.
\\

To avoid the strongly wavelength-dependent absorption, we use a graphene device that is supported by a transparent substrate that consists of 1 mm thick quartz (``transparent substrate device", see Supp.\ Info). The flake contains both single- and bilayer graphene, with PTE photovoltage generation at the interface \cite{Xu2010}. Figure 3b shows the responsivity spectrum at the monolayer-bilayer interface, together with the measured graphene absorption on a similar device. These data are obtained from spatially resolved measurements (Fig.\ 3c), which show that the spatial extent of the photoresponse does not change with wavelength. The spectral response (for constant excitation power) of the device is strikingly constant over this broad range of excitation wavelengths. The flat $R_{\rm PC}$ shows that decreasing the number of incident photons (by increasing the photon energy) does not lead to a decrease in photovoltage: thus a higher photon energy gives a larger photovoltage, as in Fig.\ 1b. This is in stark contrast with photovoltaic detectors based on semiconductors, where the photovoltage generally decreases for increasing photon energy, meaning that excess energy is lost \cite{Sze}. The flat $R_{\rm PC}$ at the monolayer-bilayer interface is therefore consistent with photo-thermoelectric current generation (which also applies to the graphene-metal interface; see Supp. Info).
\\

To understand the flat, broadband $R_{\rm PC}$ for constant absorbed power, we examine what this result means for the electron heating $\Delta T_{\rm el} = T_{\rm el} - T_0$ as a function of photon energy. First we note that the photovoltage scales linearly with power for all wavelengths (see Fig.\ 3d), which means that we are in the ``weak heating" regime where $\Delta T_{\rm el} < T_0$ and the electronic heat capacity is constant (in contrast to the ``strong heating" regime in the ultrafast experiment, where the scaling is sublinear). The reason for the small heating is that we use more than 10 times lower power and quasi-CW laser excitation with $\sim$20 ps laser pulses, which is longer than the cooling time of $\sim$1 ps (leading to a peak power that is 3 orders of magnitude smaller than for ultrafast excitation). In this ``weak heating" regime, the cooling rate is constant \cite{Graham2013}, which means that the conversion efficiency is not affected by the life time of the hot electrons. Furthermore, the Seebeck factor $S_2 - S_1$ does not change with excitation wavelength. Therefore, from the flat $R_{\rm PC}$ we conclude that the light-induced increase in electron temperature $\Delta T_{\rm el}$ at constant power is the same for all photon energies.
\\

This result enables us to assess the efficiency of the electron heating (in the ``weak heating" regime). We examine two alternative ultrafast energy relaxation pathways for photo-excited electrons and holes: carrier-carrier scattering and optical phonon emission (see Fig.\ 1a). Graphene optical phonons have an energy of $\sim$0.2 eV and therefore photoexcited carriers above 0.2 eV can in principle relax by emitting a phonon. The faster process of these two competing processes will dominate the ultrafast energy relaxation. We have determined the time scale of carrier-carrier scattering (through the photovoltage generation in the ``dual-gated device") to be $<$50 fs. The time scale of phonon emission is typically $<$150 fs \cite{Breusing2011, Lui2010}, and therefore this does not give a definite answer about which ultrafast relaxation process dominates. However, from the measured spectral responsivity we can extract the heating efficiency.
\\

We illustrate this by considering two contrasting cases (see Fig.\ 4a): \textit{i)} dominant coupling of photo-excited electrons to optical phonons, with a small fraction of the absorbed photon energy converted into hot electrons and \textit{ii)} dominant carrier-carrier scattering \cite{Winzer2010,SongPRB2013} with a large fraction of the energy converted into hot electrons. In case \textit{i}, the energy loss rate $dE/dt$ through optical phonon emission increases linearly with initial electron energy $E_i$, since it is governed by a constant electron-phonon coupling, and an energy-momentum scattering phase space that increases linearly with energy \cite{Tielrooij2013} (see Fig.\ 4a). Thus in case \textit{i}, a larger $E_i$ leads to more energy loss to phonons. On the other hand, in case \textit{ii}, the electron temperature scales linearly with $E_i$, because the energy of the photoexcited electron is fully transferred to the electron gas. Thus, the role of optical phonon emission can be measured by studying the scaling of $\Delta T_{\rm el}$ with $E_i$. This relationship, extracted from the photovoltage measurements, is shown in Fig.\ 4b, where we plot the internal quantum efficiency (IQE = ${{I_{\rm PC} E_i} \over{\Delta Q e}}$). The IQE represents the generated photovoltage, normalized by the number of absorbed photons. We find (nearly perfect) linear scaling of the IQE with $E_i$, as the linear fits go nearly trough the origin. This means that a higher photon energy corresponds to a larger photovoltage and thus to a larger $\Delta T_{\rm el}$, which is consistent with terahertz photoconductivity measurements \cite{SongPRB2013, Tielrooij2013, Jensen2014}, where a terahertz probe provided a measurement of the carrier temperature. We therefore conclude that the generated photovoltage comes from ultrafast, efficient photon-to-electron-heat conversion and an unmeasurably small loss to optical phonons.
\\

To show that carrier heating is indeed efficient, we calculate the expected temperature rise $\Delta T_{\rm el}$ and use the measured photovoltage to determine the Seebeck factor $S_2 - S_1$, which we then compare to its expected value. Details are given in the Methods section. To calculate $\Delta T_{\rm el}$ we use a simple heat equation and assume fully efficient carrier heating to find $\sim$0.17 K ($P_{\rm exc} =$ 50 $\mu$W, $T_0 =$ 300 K, $E_F = $ 250 meV, $\tau_{\rm cool} =$ 1 ps). We then use the measured photovoltage of $V_{\rm PTE} = I_{\rm PC} R \approx$ 10 $\mu$V, where $R$ is the device resistance (2 k$\Omega$).  We conclude that $S_2 - S_1 \approx$ 65 $\mu$V/K. This is very close to the maximum expected value using a charge puddle width of $\Delta =$ 80 meV, which gives 90 $\mu$V/K. Having confirmed that carrier heating is efficient, we aim to gain insight into the other factors that determine the overall energy conversion efficiency of the device. One important factor is the hot carrier lifetime, which should be long (low cooling rate) to lead to a larger photovoltage. We verify this by changing the ambient temperature. First we demonstrate that the efficiency of electron heating is independent of lattice temperature, as we obtain the same linear scaling through the origin for a lattice temperature of 40, 100 and 300 K. We then note that the overall photovoltage is larger for lower lattice temperatures. This is caused by the longer cooling time $\tau_{\rm cool}$ at low temperatures, due to a lower coupling between electrons and acoustic phonons \cite{Graham2013, Ma2014, Song2011}. The Seebeck coefficient decreases with temperature, meaning that the overall generated photovoltage is about a factor two larger \cite{Graham2013}.

\section*{Conclusion}

The unique femtosecond time resolution and the related high carrier heating efficiency are very encouraging results for bias-free (passive) PTE photodetectors. To improve the internal quantum efficiency of $\sim$1\% (Fig.\ 4b), an interesting approach (see Methods for details) would be to use ultraclean, defect-free graphene, such as in Ref.\ \cite{Wang2013}, which enables detector operation with a higher Seebeck coefficient, because the electron-hole puddle density is lower (see Methods). The small puddle width could increase the Seebeck factor $S_2 - S_1$ to $\sim$ 300 $\mu$V/K. Measuring at $E_F =$ 50 meV instead of 250 meV would furthermore lead to a larger $\Delta T_{\rm el}$, since this scales with $1/E_F$ (due to the lower heat capacity upon approaching the Dirac point). With 50 $\mu$W excitation a $\Delta T_{\rm el}$ of almost 1 K is feasible (assuming an unmodified carrier cooling rate), giving $V_{\rm PTE} \approx 300$ $\mu$V. A resistance of $R =$ 1 k$\Omega$ \cite{Wang2013} would then give an IQE of 100\% or more. Our results thus show that graphene PTE devices exhibit ultrafast, efficient and broadband photodetection. Future photodetection and light harvesting devices could exploit these graphene properties and combine them with the advantageous properties of other 2D materials. Future work could investigate the effect of the Fermi energy on the heating time, the time-resolved signal generation in devices based on the bolometric effect \cite{Freitag13b, Yan2012}, and the effect of the substrate on the heat dynamics.
\\

\section*{Methods}

\subsection*{Simulation of the photocurrent dynamics}

Here we describe how we extract the hot electron dynamics from the photocurrent dip $\Delta I_{\rm PC} (t)$. This analysis draws on the procedure described in Ref.\ \cite{Graham2013}. We start with a laser-induced change in electron temperature for single pulse excitation assuming a linear response: $\Delta T_{\rm el,1p} (t') = (1 - e^{-t' / \tau_{\rm heat}})e^{-t' / \tau_{\rm cool}}$. Here $t'$ is the 'real' time, $\tau_{\rm heat}$ the electron heating time and $\tau_{\rm cool}$ the electron cooling time. We then introduce a nonlinearity (any type of nonlinearity works) of the form $\Delta T'_{\rm el,1p} = \sqrt{1 + {{\Delta T_{\rm el,1p}}/{T_{0}}}} - 1$ (with $T_0$ the lattice temperature) and calculate the integrated photovoltage generated by two pulses well separated in time (adding up independently): $2 \cdot V_{\rm 1p} = 2 \cdot \int_0^\infty \Delta T'_{\rm el,1p}dt'$. We now follow the same procedure for two-pulse excitation where the pulses arrive with a separation of delay time $t$ and obtain $\Delta T_{\rm el,2p}$ (linear response) and
$\Delta T'_{\rm el,2p}$ (with nonlinearity). This temperature change $\Delta T'_{\rm el,2p}$ is shown in Fig.\ 1d of the main text in 'real' time for two different delay times $t$. We use these electron temperature dynamics to calculate the integrated voltage $V_{\rm 2p}$, and repeat this for a range of delay times $t$. The photocurrent dip as a function of delay time is then given by $\Delta I_{\rm PC} (t) \propto 2 \cdot V_{\rm 1p} - V_{\rm 2p} (t)$. This dip grows upon approaching $t=0$, but flattens when $t < \tau_{\rm heat}$. The reason for this is that when the two pulses arrive at exactly the same time ($t=0$), there is no heating yet due to any of the two pulses and they independently start leading to electron heating and thus contributing to the photovoltage. As a result, the photocurrent dip flattens around $t=0$. We note that the dip should not disappear, since we measure the time-integrated photocurrent and once the electrons start heating up, the heating induced by the two pulses is no longer independent. Finally, we point out that all dynamics are determined by the local (at the junction) carrier heating and PTE voltage creation. All dynamics that happen after this (propagation of the potential to the contacts, signals traveling through the cables, etc.) do no affect the results of our specific experiment. To measure transit times, one can use two-pulse excitation with two different focus spots \cite{Son2014}.
\\

\subsection*{Calculation of $\Delta T$, $V_{\rm PTE}$ and $S_2-S_1$}

We calculate the theoretical and experimental photovoltage $V_{\rm PTE}$ in the case of excitation with $P_{\rm exc} =$ 50 $\mu$W of light (of which $\sim$2 \% is absorbed, See Fig.\ 3b), in the case of the ``transparent substrate device", where the photovoltage is created at the interface of monolayer and bilayer graphene, at room temperature. We calculate the temperature increase from a simple heat equation (in the case of a cooling length that is smaller than the spot size) to be \cite{Freitag2013}:

\be \Delta T = T_{\rm el} - T_0 = {{\Delta Q } \over {2 d^2 \sqrt{4 \pi \ln 2} \cdot \Gamma_{\rm cool}}}  \approx {\rm 0.17 \hspace{0.8 mm} K} \hspace{1mm} .\ee

Here, the laser focus spot size is $d \approx$ 1.5 $\mu$m (see Fig.\ 3c), and the cooling rate is given by $\Gamma_{\rm cool} = {{\alpha T_{\rm el}} \over {\tau_{\rm cool}}}$, where $\alpha = {{2 \pi E_F}\over{3 \hbar^2 v_F^2}} k_B^2$, with $\hbar$, $v_F$ and $k_B$ the reduced Planck constant, the Fermi velocity and Boltzmann's constant, respectively. We use $E_F = $ 0.25 eV and a cooling time of 1 ps, which we obtained from ultrafast photocurrent measurements on the same device. This yields a cooling rate of 0.4 MW/m$^2$K, close to theoretical estimates that produce 0.5-5 MW/m$^2$K \cite{Freitag2013}. To obtain the photovoltage we use \cite{Freitag2013}:

\be V_{\rm PTE} = I_{\rm PC} R \approx {\rm 10}  \hspace{0.8mm} \mu{\rm V} \hspace{1mm} ,\ee

where we use a photocurrent of 5 nA for $P_{\rm exc} =$ 50 $\mu$W and a device resistance of $R$ = 2 k$\Omega$. Using $V_{\rm PTE} = \Delta T_{\rm el} (S_2 - S_1)$, we thus obtain $(S_2 - S_1) =$ 65 $\mu$V/K. The maximum Seebeck factor is given by $(S_2-S_1)_{\rm max} = {{\pi^2 k_B^2 T}\over{3 e \Delta}} $ \cite{Gabor2011}, which gives 90 $\mu$V/K at room temperature for a charge puddle width of $\Delta =$ 80 meV.
\\

\subsection*{Acknowledgements}
We thank Justin Song, Leonid Levitov and Daan Brinks for useful discussions. KJT thanks NWO for a Rubicon fellowship. LP acknowledges financial support from Marie-Curie International Fellowship COFUND and ICFOnest program. FK acknowledges support by the Fundacio Cellex Barcelona, the ERC Career integration grant 294056 (GRANOP), the ERC starting grant 307806 (CarbonLight) and support by the E.\ C.\ under Graphene Flagship (contract no. CNECT-ICT-604391). NvH acknowledges support from ERC advanced grant ERC247330. QM and PJH have been supported by AFOSR Grant No. FA9550-11-1-0225 and a Packard Fellowship. This work made use of the Materials Research Science and Engineering Center Shared Experimental Facilities supported by the National Science Foundation (NSF) (Grant No. DMR-0819762) and of Harvard’s Center for Nanoscale Systems, supported by the NSF (Grant No. ECS-0335765).
\\

\onecolumngrid
\clearpage

\section*{Figures}

\begin{figure} [h!!!!!]
   \centering
   \includegraphics [scale=0.85]
   {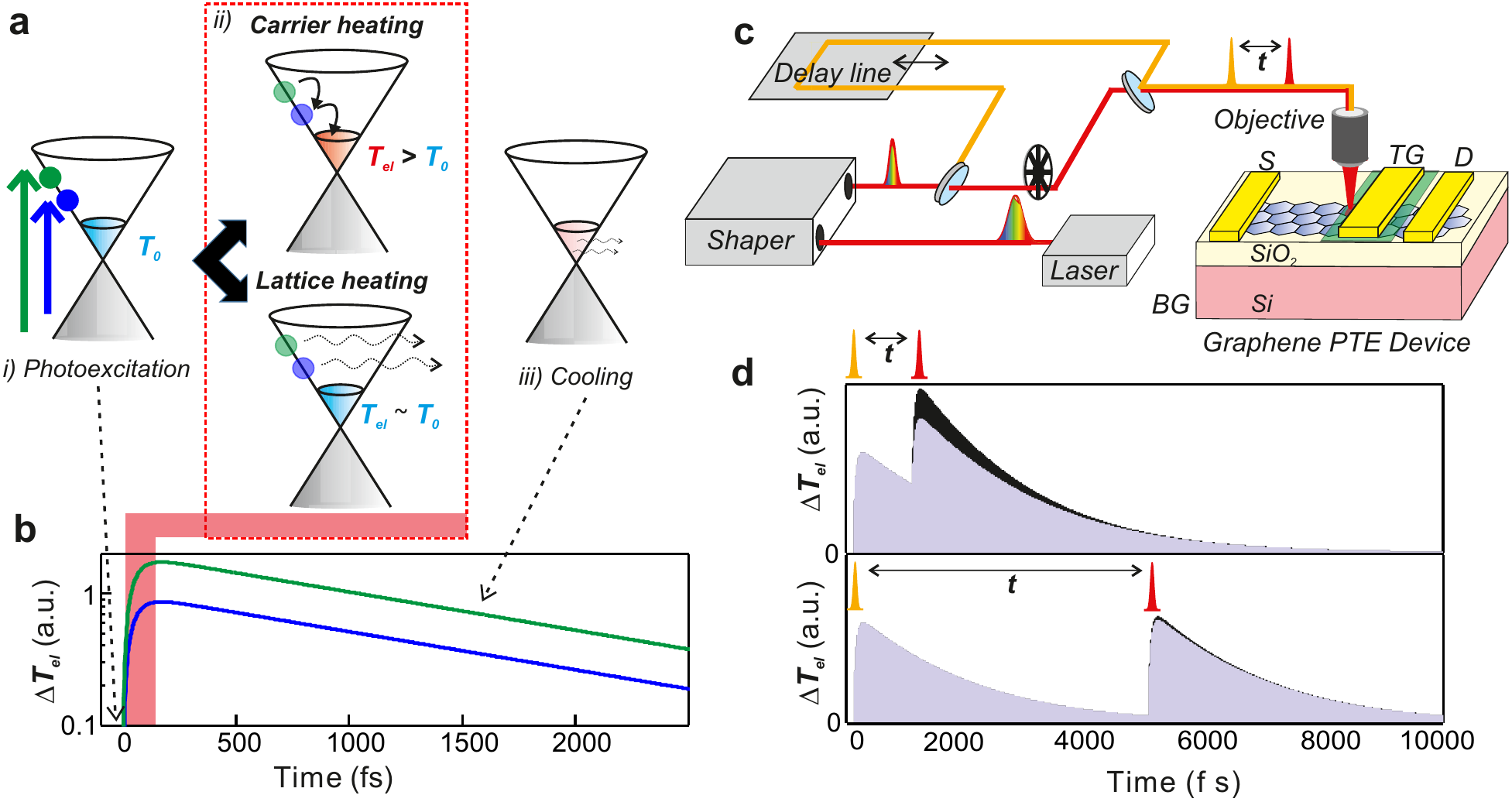}
\caption{\textbf{Hot electron dynamics and their experimental extraction.}
\textbf{a)} Schematic representation of the electron dynamics in graphene after photoexcitation with two different photon energies (green and blue arrow). After photoexcitation and electron-hole pair generation, electron heating and lattice heating take place. The former occurs through carrier-carrier scattering and leads to hot electrons \cite{Tielrooij2013, SongPRB2013, Lui2010, Gierz2013, Johannsen2013, Graham2013}, which drive a photo-thermoelectric current \cite{Gabor2011, Song2011, Xu2010}. The latter occurs through phonon emission and leads to the generation of a much smaller photovoltage, since the heat capacity of the phonon bath is much larger than that of the electron bath \cite{Song2011}.
\textbf{b)} The hot electrons lead to a local photovoltage through the photo-thermoelectric effect, with dynamics governed by the hot electron dynamics. If electron heating is efficient, a higher photon energy leads to a larger photovoltage (green line), compared to a lower photon energy (blue line).
\textbf{c)} Schematic representation of the ultrafast photovoltage setup and the ``dual-gated graphene device" (more details in the Supp.\ Info). The setup produces two ultrashort pulses, separated in time by a controllable delay time $t$. A pulse shaper compensates for pulse stretching in the objective. The compressed pulses are focused onto the device, which contains a back gate (BG) and a top gate (TG) for creation of a $pn$-junction at the interface between graphene regions of opposite doping. We read out the photocurrent through the source (S) and drain (D) contacts, using a pre-amplifier and a lock-in amplifier, synchronised with the optical chopper.
\textbf{d)} The development of the electron temperature after excitation with an ultrafast pulse pair with $t = 1$ ps in the situation of independent heating due to the two pulses (black area) and heating where the heating due to the second pulse depends on the heating due to the first pulse (purple area). The measured photocurrent $I_{\rm PC}$ is proportional to the purple area; the photocurrent dip $\Delta I_{\rm PC}$ is proportional to the difference between the black and the purple area. If $t=5$ ps the black and purple areas are very similar, i.e.\ $\Delta I_{\rm PC} \approx 0$. Thus, the photocurrent dip as a function of $t$ reflects the heating dynamics.
}
    \label{Fig_ConceptHeatingCooling}
\end{figure}

\clearpage

\begin{figure} [h!!!!!]
   \centering
   \includegraphics [scale=0.75]
   {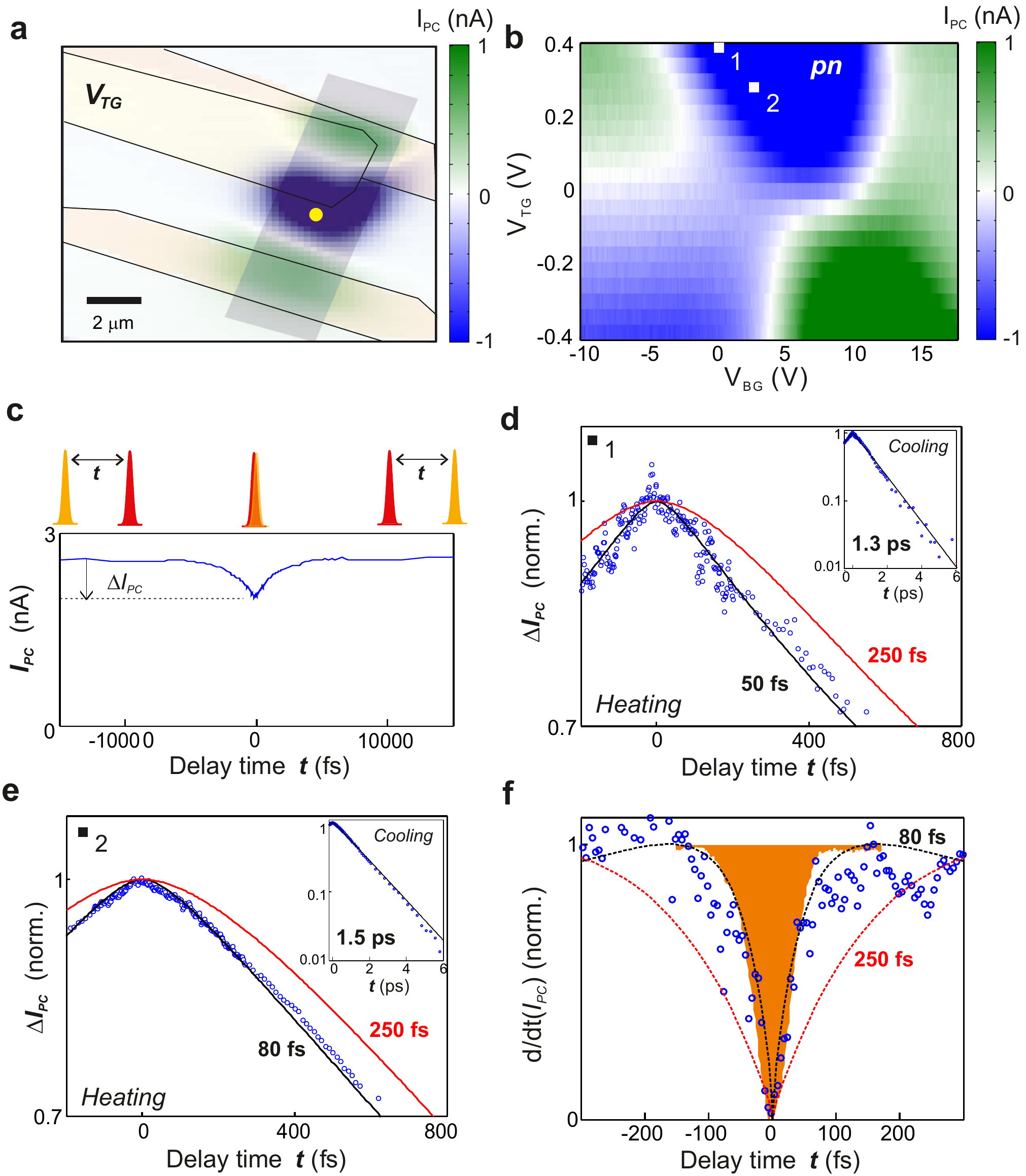}
\caption{\textbf{Femtosecond sensing of hot electrons.}
\textbf{a)} Scanning photocurrent image (green-blue color scale) of the ``dual-gated device", with indications of the edges of the two contacts and the topgate, and the location of the graphene flake (grey rectangle). At the edge of the topgate, there is an interface of two graphene regions whose Fermi energy is separately controlled by the voltages on the back gate and the top gate, here tuned to the $pn$-junction regime.
\textbf{b)} The photocurrent as a function of gate voltages at the position of the yellow dot in panel \textbf{a}, showing clear multiple sign reversals that are indicative of PTE current generation \cite{Song2011, Gabor2011}.
\textbf{c)} The photovoltage as a function of delay time $t$ shows a dip $\Delta I_{\rm PC}$ when the two pulses overlap and recovers with dynamics that correpond to the hot electron dynamics. We show the dynamics of the (normalized) time-resolved photovcurrent dip $\Delta I_{\rm PC}$ around $t=0$ (blue circles) for gate configurations 1 \textbf{(d)} and 2 \textbf{(e)}, both in the $pn$-junction regime. The black solid lines describe the model results (see Methods), using a heating time of 50 (80) fs and a cooling time of 1.3 (1.5) ps for gate configuration 1 (2). The red lines show the modelled dip with a slower heating time scale of 250 fs, which is incompatible with the data. The insets show the data and model results over a larger time range.
\textbf{f)} The time derivative of the photocurrent dip (blue circles) in gate configuration 2, together with the derivative of the modelled photocurrent dip with heating time scales of 80 (black dashed line) and 250 fs (red dashed line). The orange area represents the optically measured cross correlate using a nonlinear crystal.
}
    \label{Fig_UltrafastSetup}
\end{figure}

\clearpage

\begin{figure} [h!!!!!]
   \centering
   \includegraphics [scale=0.8]
   {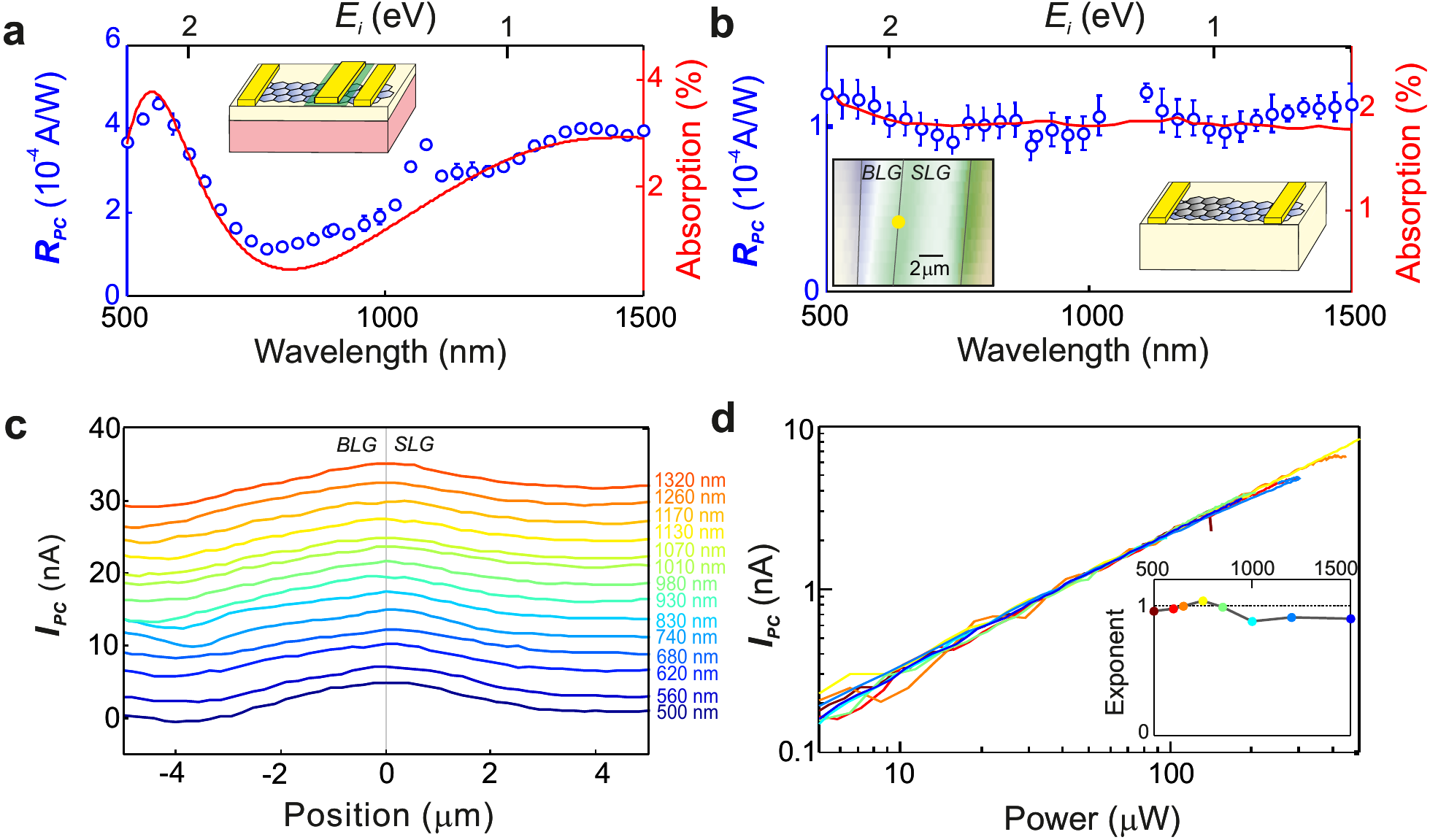}
\caption{\textbf{Spectral response.} \textbf{a)} Responsivity as a function of excitation wavelength (blue data points, left axis) and modelled absorption spectrum (red line, right axis) for the ``dual-gated device" in the $pn$-configuration, measured with a fixed power. The inset shows the device layout, with a 300 nm SiO$_2$ substrate on top of Si. Error bars are calculated from independent measurement scans and represent the 68\% confidence interval.
\textbf{b)} Responsivity as a function of excitation wavelength (blue data points, left axis) and modelled absorption spectrum (red line, right axis) for the ``transparent substrate device" at the monolayer-bilayer graphene interface ($\sim$50 $\mu$W power). Error bars are calculated from independent measurement scans and represent the 68\% confidence interval. The left inset shows a scanning photovoltage image with photovoltage generation at the monolayer-bilayer (SLG-BLG) interface. The right inset shows the device layout, with a 1 mm thick SiO$_2$ substrate. The agreement between the responsivity and absorption curve shows that a lower number of absorbed photons (shorter wavelength, higher photon energy) does not lead to a lower responsivity, consistent with PTE current generation \cite{Song2011}. \textbf{c)} Photocurrent as a function of laser spot position for wavelengths between 500 and 1500 nm (offset for clarity), while scanning the laser through the interface between monolayer and bilayer graphene, showing constant photovoltage amplitudes and spatial extent. The laser focus spot size is 1.5 $\pm$ 0.15 $\mu$m. \textbf{d)} Power dependence of the photovoltage with the laser focused at the SLG-BLG interface (see inset in panel \textbf{b}) for a range of wavelengths (see inset for the wavelength corresponding to each color), showing linear scaling, which corresponds to the ``weak heating" regime, where $\Delta T_{\rm el} < T_0$. The inset shows the fitted power exponent, which is close to one.
}
    \label{Fig_UltrafastResults}
\end{figure}

\clearpage

\begin{figure} [h!!!!!]
   \centering
   \includegraphics [scale=0.8]
   {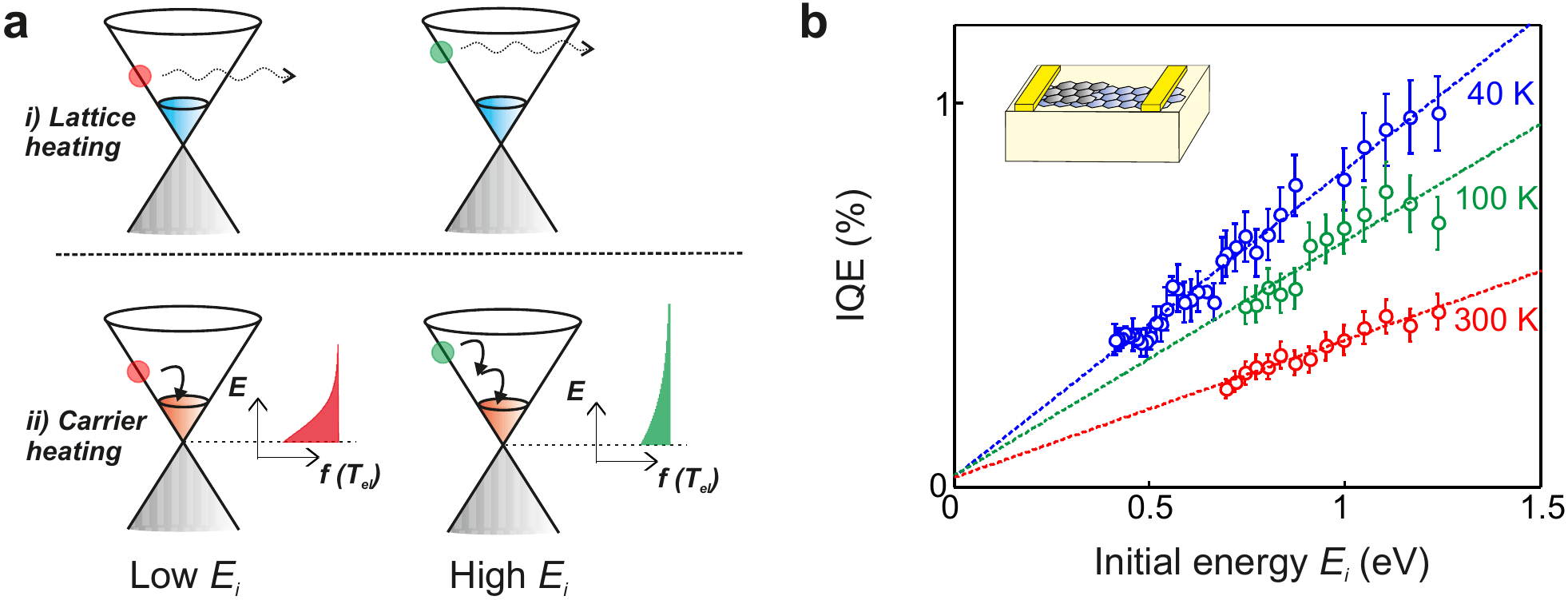}
\caption{\textbf{Electron heating efficiency.} \textbf{a)} Schematic representation of ultrafast energy relaxation after photoexcitation, through phonon emission, which leads to lattice heating (top), and through carrier-carrier scattering, which leads to carrier heating (bottom). In the case of lattice heating, a larger photon energy leads to a larger phase space to scatter to and therefore more energy is transferred to the phonon bath, predicting sub-linear scaling of the photovoltage (normalized by absorbed photon density) with photon energy. In the case of carrier heating, a larger photon energy leads to a hotter electron distribution (see the smeared Fermi-Dirac distributions next to the Dirac cones), predicting linear scaling of photocurrent (normalized by absorbed photon density) with photon energy (in the weak heating regime). \textbf{b)} The internal quantum efficiency (IQE), which represents the photovoltage normalized by absorbed photon density, as a function of initial electron energy after photoexcitation for ambient temperatures $T_0 =$ 40, 100 and 300 K. Error bars are calculated from independent measurement scans and represent the 68\% confidence interval. The linear scaling through the origin shows that heating dominates the ultrafast energy relaxation and  therefore that electron heating is efficient. }
    \label{Fig_LinearScaling}
\end{figure}

\end{document}